\documentclass[preprint,showpacs,preprintnumbers,amsmath,amssymb,floatfix]{revtex4}
\usepackage{amsmath}

\usepackage{graphicx}
\usepackage{pdfpages}
\usepackage{dcolumn}
\usepackage{bm}
\usepackage{amsmath} 
\usepackage{amsfonts}
\usepackage{amssymb}
\usepackage{url}
\usepackage{microtype}
\usepackage{graphicx}
\usepackage{appendix}
\usepackage{esvect}%

\newcommand{\qed}{\nobreak \ifvmode \relax \else
      \ifdim\lastskip<1.5em \hskip-\lastskip
      \hskip1.5em plus0em minus0.5em \fi \nobreak
      \vrule height0.75em width0.5em depth0.25em\fi}

\begin{document}

\preprint{}
\title{8 Boolean Atoms Spanning the 256-Dimensional Entanglement-Probability Three-Set Algebra of the Two-Qutrit Hiesmayr-L{\"o}ffler  Magic Simplex of Bell States}
\author{Paul B. Slater}
 \email{paulslater@ucsb.edu}
\affiliation{%
Kavli Institute for Theoretical Physics, University of California, Santa Barbara, CA 93106-4030\\
}
\date{\today}
            
\begin{abstract}
We obtain formulas (bot. p. 12)--including $\frac{2}{121}$ and $\frac{4 \left(242 \sqrt{3} \pi -1311\right)}{9801}$--for the  eight atoms (Fig.~\ref{fig:Venn}), summing to 1, which span a 256-dimensional three-set (P, S, PPT) entanglement-probability boolean algebra for the two-qutrit Hiesmayr-L{\"o}ffler  states. PPT denotes positive partial transpose, while P and S  provide the Li-Qiao necessary {\it and} sufficient conditions for entanglement.  The constraints ensuring entanglement are $s> \frac{16}{9} \approx 1.7777$ and $p> \frac{2^{27}}{3^{18} \cdot 7^{15} \cdot13} \approx 5.61324 \cdot 10^{-15}$. Here, 
$s$ is the square of the  sum (Ky Fan norm) of the eight singular values  of the $8 \times 8$ correlation matrix in the  Bloch representation, and $p$, the square of the product of the singular values. In the two-{\it ququart} Hiesmayr-L{\"o}ffler  case, one constraint is $s>\frac{9}{4} \approx 2.25$, while  $\frac{3^{24}}{2^{134}} \approx 1.2968528306 \cdot 10^{-29}$ is an upper bound on the appropriate $p$ value, with an entanglement probability $\approx 0.607698$.
The $S$ constraints, in both cases, prove equivalent to the well-known CCNR/realignment criteria. Further, we  detect and verify--using software of A. Mandilara--pseudo-one-copy undistillable (POCU) negative partial transposed two-qutrit states distributed over the surface of the separable states. Additionally, we study the {\it best separable approximation} problem within this two-qutrit setting, and obtain explicit decompositions of separable states into the sum of eleven product states. Numerous quantities of interest--including the eight atoms--were, first, estimated using
a quasirandom procedure. 
\end{abstract}
 

\maketitle
\section{Introduction}
In our recent preprint ``Jagged Islands of Bound Entanglement and Witness-Parameterized Probabilities'' \cite{slater2019bound}, we reported a PPT (positive partial transpose) Hilbert-Schmidt probability of $\frac{8 \pi }{27 \sqrt{3}} \approx 0.537422$ for the Hiesmayr-L{\"o}ffler two-qutrit magic simplex of Bell states (and $\frac{1}{2}+\frac{\log \left(2-\sqrt{3}\right)}{8 \sqrt{3}} \approx 0.404957$ for the two-ququart counterpart) \cite{hiesmayr2014mutually}.
Additionally, we  utilized their  mutually unbiased bases (MUB) test and  the Choi $W^{(+)}$ witness test \cite{ha2011one,chruscinski2018generalizing},  obtaining a total entanglement (that is, bound plus  ``non-bound"/``free") probability for  each test of $\frac{1}{6} \approx 0.16667$,  while their union and intersection $\frac{2}{9} \approx 0.22222$ and  $\frac{1}{9} \approx 0.11111$, respectively. 
The same bound-entangled probability  $-\frac{4}{9}+\frac{4 \pi }{27 \sqrt{3}}+\frac{\log (3)}{6} \approx 0.00736862$ was achieved with each witness--the sets (``jagged islands") detected, however, having void intersection.  The results were summarized 
in Table~\ref{tab:Main} there, repeated here. 
\begin{table} 
\begin{center}
 \begin{tabular}{||c c c ||} 
 \hline
 Set & Probability  & Numerical Value  \\ 
 \hline
 \hline
 -------& 1 & 1. \\ 
 \hline
 PPT & $\frac{8 \pi }{27 \sqrt{3}}$ & 0.537422  \\
 \hline
 MUB & $\frac{1}{6}$ & 0.1666667  \\
 \hline
Choi & $\frac{1}{6}$ & 0.1666667  \\
 \hline
 $\text{PPT}\land \text{MUB}$ & $-\frac{4}{9}+\frac{4 \pi }{27 \sqrt{3}}+\frac{\log (3)}{6}$ & 0.00736862 \\
 \hline
 $\text{PPT}\land \text{Choi}$ & $-\frac{4}{9}+\frac{4 \pi }{27 \sqrt{3}}+\frac{\log (3)}{6}$ & 0.00736862 \\
 \hline
 $\text{MUB}\land \text{Choi}$ & $\frac{1}{9}$ & 0.11111 \\
 \hline
 $\text{MUB}\lor \text{Choi}$ & $\frac{2}{9}$ & 0.22222 \\
  \hline
   $\neg \text{MUB}\land \text{Choi}$ & $\frac{1}{18}$ & 0.05555 \\
    \hline
   $\text{MUB}\land  \neg \text{Choi}$ & $\frac{1}{18}$ & 0.05555 \\
 \hline
 $\text{PPT}\land \neg \text{MUB}$ & $\frac{1}{162} \left(72+8 \sqrt{3} \pi -27 \log (3)\right)$ & 0.5300534 \\
 \hline
 $\text{PPT}\land \neg \text{Choi}$ & $\frac{1}{162} \left(72+8 \sqrt{3} \pi -27 \log (3)\right)$ & 0.5300534 \\
 \hline
$\text{PPT}\land \text{MUB}\land \text{Choi}$ & 0 & 0 \\
\hline
$\text{PPT}\land(\text{MUB}\lor \text{Choi})$ & $-\frac{8}{9}+\frac{8 \pi }{27 \sqrt{3}}+\frac{\log (3)}{3}$ & 0.0147372 \\
\hline
$\neg \text{PPT}\land \text{MUB}$ & $\frac{1}{3}+\frac{22518 \sqrt{3}}{91}+\frac{3888 \sqrt{3}}{7 \pi }-\frac{10939 \pi }{27
   \sqrt{3}}-\frac{\log (3)}{8}$ & 0.1592980 \\
\hline
$\neg \text{PPT}\land \text{Choi}$ & $\frac{1}{3}+\frac{22518 \sqrt{3}}{91}+\frac{3888 \sqrt{3}}{7 \pi }-\frac{10939 \pi }{27
   \sqrt{3}}-\frac{\log (3)}{8}$ & 0.1592980 \\
  \hline
 $\neg \text{PPT}\land \neg \text{MUB}$ &$\frac{1}{162} \left(9 (7+\log (27))-8 \sqrt{3} \pi \right)$ & 0.303279920
 \\
 \hline
  \hline
 $\neg \text{PPT}\land \neg \text{Choi}$ &$\frac{1}{162} \left(9 (7+\log (27))-8 \sqrt{3} \pi \right)$ & 0.303279920
 \\
 \hline
 $\neg \text{PPT}\land \neg \text{MUB} \land \neg \text{Choi}$ & $\frac{1}{9} (3 \log (3)-1)$ & 0.255092985 \\
  \hline
 $ \text{PPT}\land \neg \text{MUB} \land \neg \text{Choi}$ & $\frac{1}{9} (8-3 \log (3))$ & 0.5226847927 \\
 \hline
 $\text{PPT}\lor (\text{MUB}\land \text{Choi})$ & $\frac{1}{81} \left(9+8 \sqrt{3} \pi \right)$ & 0.648533145 \\
 \hline
\end{tabular}
\caption{\label{tab:Main}Various Hilbert-Schmidt probabilities for the Hiesmayr-L{\"o}ffler $d=3$ two-qutrit model. Notationally, $\neg$ is the negation logic operator (NOT); $\land$ is the conjunction logic operator (AND); and $\lor$  is the disjunction logic operator (OR). The mutually unbiased and Choi witness tests are indicated.}
\end{center}
\end{table}
(We will supplement these results in Table~\ref{tab:Second} below, presenting formulas--our major advance--for the titular 8 boolean atoms.)

Further, application there of the realignment (CCNR) test for entanglement \cite{chen2002matrix,shang2018enhanced} yielded an entanglement probability  of $\frac{1}{81} \left(27+\sqrt{3} \log \left(97+56 \sqrt{3}\right)\right) \approx 0.445977$ and an exact bound-entangled probability of $\frac{2}{81} \left(4 \sqrt{3} \pi -21\right) \approx 0.0189305$. (Thus, as we will find through independent means, the entanglement probability attributable to the Li-Qiao [sum] $S$ constraint \cite{li2018necessary,li2018separable}--but ignoring their 
[product] $P$ constraint--equals $(1-\frac{8 \pi }{27 \sqrt{3}})+\frac{2}{81} \left(4 \sqrt{3} \pi -21\right) = \frac{13}{27} \approx 0.481481$. However, in the original arXiv posting of this paper, we reported $\frac{13}{27}$ as the {\it entire} entanglement probability--but now must revise it to $1-\frac{21}{44}=\frac{23}{44} \approx 0.522727$.) In the two-ququart Hiesmayr-L{\"o}ffler case, the analogous target entanglement probability appears to be $(1-(\frac{1}{2}+\frac{\log \left(2-\sqrt{3}\right)}{8 \sqrt{3}})) + 0.012654 \approx 0.607698$.

Also, in a pair of recent reprints ``Archipelagos of Total Bound and Free Entanglement'' \cite{slater2020archipelagos} and ``Archipelagos of Total Bound and Free Entanglement. II'' \cite{slater2020archipelagos2}, we implemented  the necessary {\it and} sufficient conditions recently put forth by Li and Qiao \cite{li2018necessary,li2018separable} (cf. \cite{peled2020correlation}) for the three-parameter qubit-ququart model,
\begin{equation} \label{rhoAB}
\rho_{AB}^{(1)}=\frac{1}{2 \cdot 4} \textbf{1} \otimes \textbf{1} +\frac{1}{4} (t_1 \sigma_1 \otimes \lambda_1+t_2 \sigma_2 \otimes \lambda_{13}+t_3 \sigma_3 \otimes \lambda_3),
\end{equation}
where $t_{\mu} \neq 0$, $t_{\mu} \in \mathbb{R}$, and $\sigma_i$ and $\lambda_{\nu}$ are SU(2) (Pauli matrix) and SU(4) generators, respectively (cf. \cite{singh2019experimental}).
We also examined there, certain three-parameter two-ququart and two-qutrit scenarios.
\section{Li-Qiao Hiesmayr-L{\"o}ffler Two-Qutrit Analyses}
Here, we seek--in two different manners--to extend these procedures developed by Li and Qiao to the Hiesmayr-L{\"o}ffler two-qutrit magic simplex of Bell states \cite{hiesmayr2014mutually}, earlier studied by us in \cite{slater2019bound}. To do so, constitutes a substantial challenge, since now the associated correlation matrix 
of the Bloch representation of the bipartite state 
\begin{equation} \label{HLCorrelation}
\rho_{HL}= 
\frac{1}{9} \textbf{1} \otimes \textbf{1} + \frac{1}{4} \Big(t_9 \lambda_{3} \otimes \lambda_{8} +t_{10} \lambda_{8} \otimes \lambda_{3} +\Sigma_{i=1}^8 t_i \lambda_{i} \otimes \lambda_{i})\Big) .    
\end{equation}
is $8 \times 8$, rather than $2 \times 2$ or $3 \times 3$ as in our previous studies  and those of Li and Qiao.
(Interestingly, in the three-dimensional matrix [Gell-mann] representation of $SU(3)$, the Cartan subalgebra is the set of linear combinations [with real coefficients] of the two matrices 
$\lambda_3$ and $\lambda_8$, which commute with each other.)
In the simplifying parameterization of the Hiesmayr--L{\"o}ffler states introduced in \cite[sec. II.A]{slater2019bound}, 
\begin{equation} \label{d=3HL}
\rho_{HL}=\left(
\begin{array}{ccccccccc}
 \gamma _1 & 0 & 0 & 0 & \gamma _2 & 0 & 0 & 0 & \gamma _2 \\
 0 & Q_2 & 0 & 0 & 0 & 0 & 0 & 0 & 0 \\
 0 & 0 & \gamma _3 & 0 & 0 & 0 & 0 & 0 & 0 \\
 0 & 0 & 0 & \gamma _3 & 0 & 0 & 0 & 0 & 0 \\
 \gamma _2 & 0 & 0 & 0 & \gamma _1 & 0 & 0 & 0 & \gamma _2 \\
 0 & 0 & 0 & 0 & 0 & Q_2 & 0 & 0 & 0 \\
 0 & 0 & 0 & 0 & 0 & 0 & Q_2 & 0 & 0 \\
 0 & 0 & 0 & 0 & 0 & 0 & 0 & \gamma _3 & 0 \\
 \gamma _2 & 0 & 0 & 0 & \gamma _2 & 0 & 0 & 0 & \gamma _1 \\
\end{array}
\right) ,   
\end{equation}
where $\gamma_1=\frac{1}{3} \left(Q_1+2 Q_3\right),\gamma_2=\frac{1}{3} \left(Q_1-Q_3\right)$, and $\gamma_3=\frac{1}{3} \left(-Q_1-3 Q_2-2 Q_3+1\right)$,
we have $t_1=t_4=t_6=\frac{2}{3} \left(Q_1-Q_3\right), t_2=t_5=-\frac{2}{3} \left(Q_1-Q_3\right), t_3=t_8 =-(\frac{1}{3}+Q_1+2 Q_3)$. Further, $t_9=\frac{Q_1+6 Q_2+2 Q_3-1}{\sqrt{3}}$ and $t_{10}=-t_9$.

The requirement that $\rho_{HL}$ be a nonnegative definite density matrix--ensured by requiring that its nine leading nested minors all be nonnegative \cite{prussing1986principal}--takes the form \cite[eqs. (29)]{slater2019bound},
\begin{equation} \label{constraint1}
Q_1>0\land Q_2>0\land Q_3>0\land Q_1+3 Q_2+2 Q_3<1.   
\end{equation}
Additionally, the constraint that the partial transpose of the $9 \times 9$ density matrix be  nonnegative definite is
\begin{equation} \label{d=3PPT}
 Q_1>0\land Q_3>0\land Q_1+3 Q_2+2 Q_3<1\land Q_1^2+3 Q_2 Q_1+\left(3
   Q_2+Q_3\right){}^2<3 Q_2+2 Q_1 Q_3.   
\end{equation}
Further, the Hiesmayr-L{\"o}ffler  mutually-unbiased-bases (MUB) criterion for bound entanglement, $I_4 = \Sigma_{k=1}^4 
C_{A_k,B_k} >2$, where $C_{A_k,B_k}$ are correlation functions for observables $A_k,B_k$ \cite[Fig. 1]{hiesmayr2013complementarity} is
\begin{equation} \label{MUBconstraint}
 Q_1>3 Q_2+4 Q_3 , 
\end{equation}
In the Hiesmayr-L{\"o}ffler $d=3$ two-qutrit 
density-matrix setting, the Choi-witness entanglement requirement that $\mbox{Tr} [W \rho_{HL}]<0$ assumes the  form 
\begin{equation} \label{Choi}
2 Q_3+1-2 Q_1-3 Q_2 <0.    
\end{equation}
The realignment constraint that, if satisfied, ensures entanglement is 
\begin{equation} \label{realignment}
 \sqrt{-9 Q_2-6 Q_3+3 \left(Q_1^2+\left(3 Q_2+4 Q_3-1\right) Q_1+9 Q_2^2+4
   Q_3^2+6 Q_2 Q_3\right)+1}+
\end{equation}
\begin{displaymath}
+3 |Q_1-Q_3| >1.
\end{displaymath}
\subsection{Singular values  of the $8 \times 8$ Hiesmayr--L{\"o}ffler two-qutrit  correlation matrix}
The pair ($P,S$) of entanglement constraints in the Li-Qiao framework, for which we seek the appropriate bounds,  would be based on the eight singular values  of the $8 \times 8$ correlation matrix for the Hiesmayr--L{\"o}ffler model--bipartite in nature--under examination. 
 (We should note that the correlation matrix for this two-qutrit model is non-diagonal in nature, since there are terms in  the expansion (\ref{HLCorrelation}) of the form $\lambda_3 \otimes \lambda_8$ and $\lambda_8 \otimes \lambda_3$. The coefficients of these terms in the indicated reparameterization being $\frac{Q_1+6 Q_2+2 Q_3-1}{\sqrt{3}}$ and $-\frac{Q_1+6 Q_2+2 Q_3-1}{\sqrt{3}}$, respectively, as noted earlier.) 
 
 Entanglement is achieved if {\it either} the square ($p$) of the {\it product} of the eight singular values exceeds a certain threshold, {\it or} the square ($s$) of the {\it sum} (the Ky Fan norm) of   the singular values exceeds a corresponding threshold.
 Our research here is first focused on determining the appropriate thresholds to employ. (The set of two-qutrit states 
 satisfying the first [product-form] of these two constraints we denote $P$ and the second [sum-form], $S$.)
 
 To so proceed, we found that six of the eight singular values of the correlation matrix of (\ref{HLCorrelation}) are $\frac{2}{3} \sqrt{\left(Q_1-Q_3\right){}^2}$ and the remaining two are $\frac{2}{3} \sqrt{-9 Q_2-6 Q_3+3 \left(Q_1^2+\left(3 Q_2+4 Q_3-1\right) Q_1+9 Q_2^2+4 Q_3^2+6 Q_2 Q_3\right)+1}$. The square of the product of the eight values is, then,
\begin{equation} \label{MultiplicativeNorm}
 \frac{65536 \left(Q_1-Q_3\right){}^{12} \left(3 Q_1^2+3 \left(3 Q_2+4 Q_3-1\right) Q_1+27
   Q_2^2+9 Q_2 \left(2 Q_3-1\right)+6 Q_3 \left(2 Q_3-1\right)+1\right){}^2}{43046721}   
\end{equation}
and the square  of their sum  is 
\begin{equation}\label{AdditiveNorm}
 \left(\frac{4 \sqrt{\zeta }}{3}+4 \sqrt{\left(Q_1-Q_3\right){}^2}\right){}^2  
\end{equation}
where (cf. (\ref{realignment}))
\begin{equation}
 \zeta= -9 Q_2-6 Q_3+3 \left(Q_1^2+\left(3 Q_2+4 Q_3-1\right) Q_1+9 Q_2^2+4 Q_3^2+6 Q_2
   Q_3\right)+1 .
\end{equation}
These are the two quantities--in the Li-Qiao framework--for which we must find suitable  lower bounds. If a particular Hiesmayr-L{\"o}ffler state exceeds either bound it is necessarily entangled.
We, preliminarily, found that the maxima--over the entire magic simplex (of both entangled and separable states)--for $P$  is   $\frac{65536}{43046721} =(\frac{2}{3})^{16} \approx 0.00152$ and for $S$, $ \frac{256}{9} \approx 28.4444$. But, we principally desire the maxima over solely the separable states--since delineating such states is, in general, intrinsically difficult 
\cite{gurvits2003classical}.

Thus, we now restrict the search for the maxima to the Hiesmayr--L{\"o}ffler states with positive partial transpose, but which are {\it not} bound-entangled according to the realignment test.
Then, our numerics indicated that the maxima are   $\frac{134217728}{23910933822616040487651}=\frac{2^{27}}{3^{18} \cdot 7^{15} \cdot13} \approx 5.61324 \cdot 10^{-15}$
 \cite{DenominatorQuestion} for $P$  and $\frac{16}{9} \approx 1.7777$ for $S$ (at $Q_1=\frac{1}{3}, Q_2=0, Q_3=\frac{1}{3}$). (This last maximum can also be achieved at  $Q_1=\frac{1}{4}, Q_2=\frac{1}{24} (3 -\sqrt{5}), Q_3=0$--which in the original Hiesmayr-L{\"o}ffler coordinates, converts to $q_1 = \frac{5}{24} \left(\sqrt{5}-3\right),q_2=
   -1-\frac{\sqrt{5}}{3},q_3=-\frac{\sqrt{5}}{4}$. If, on the other hand, we simply search for the maxima over the Hiesmayr--L{\"o}ffler two-qutri states with positive partial transpose--within which all the separable states must lie, but now do {\it not} omit those states that are bound-entangled based on the realignment test, we  obtain the larger values for $S$, $s=\frac{25}{9} \approx 2.7777$, and $p=\frac{2^{28}}{3^{16} \cdot 7^{14}} \approx 9.194481490 \cdot 10^{-12}$ for $P$ (at $Q_1=\frac{2}{7}, Q_2=\frac{4}{21}, Q_3=0$).)

Enforcement of  the constraint $S \equiv s>\frac{16}{9}$ proves, interestingly (algebraically demonstrable),  fully equivalent (at least for $d=3$) to the application of both the realignment (CCNR)  and SIC POVMs   tests  \cite{chen2002matrix,shang2018enhanced}, in yielding a total entanglement probability  of $\frac{1}{81} \left(27+\sqrt{3} \log \left(97+56 \sqrt{3}\right)\right) \approx 0.445977$ and a bound-entanglement probability of $\frac{2}{81} \left(4 \sqrt{3} \pi -21\right) \approx 0.0189035$. (The realignment bound-entangled ``island" completely contains the corresponding Choi and MUB islands, with an additional probability of $\frac{1}{27} (10-9 \log (3)) \approx 0.00416627$ \cite[Fig. 25]{slater2019bound}. The bound $s \leq \frac{16}{9}$ is one of the known results for separability, using the Bloch representation \cite[eq. (48)]{li2018separable}.) 

We have also, interestingly, found that of this bound-entanglement probability of $\frac{2}{81} \left(4 \sqrt{3} \pi -21\right) \approx 0.0189035$, the measure $\frac{2}{121}  \approx 0.0165289$ is also yielded by the $P \equiv p>\frac{134217728}{23910933822616040487651}=\frac{2^{27}}{3^{18} \cdot 7^{15} \cdot13} \approx 5.61324 \cdot 10^{-15}$ constraint.
\subsection{Graphic representations} \label{Graphic}
\begin{figure}
    \centering
    \includegraphics{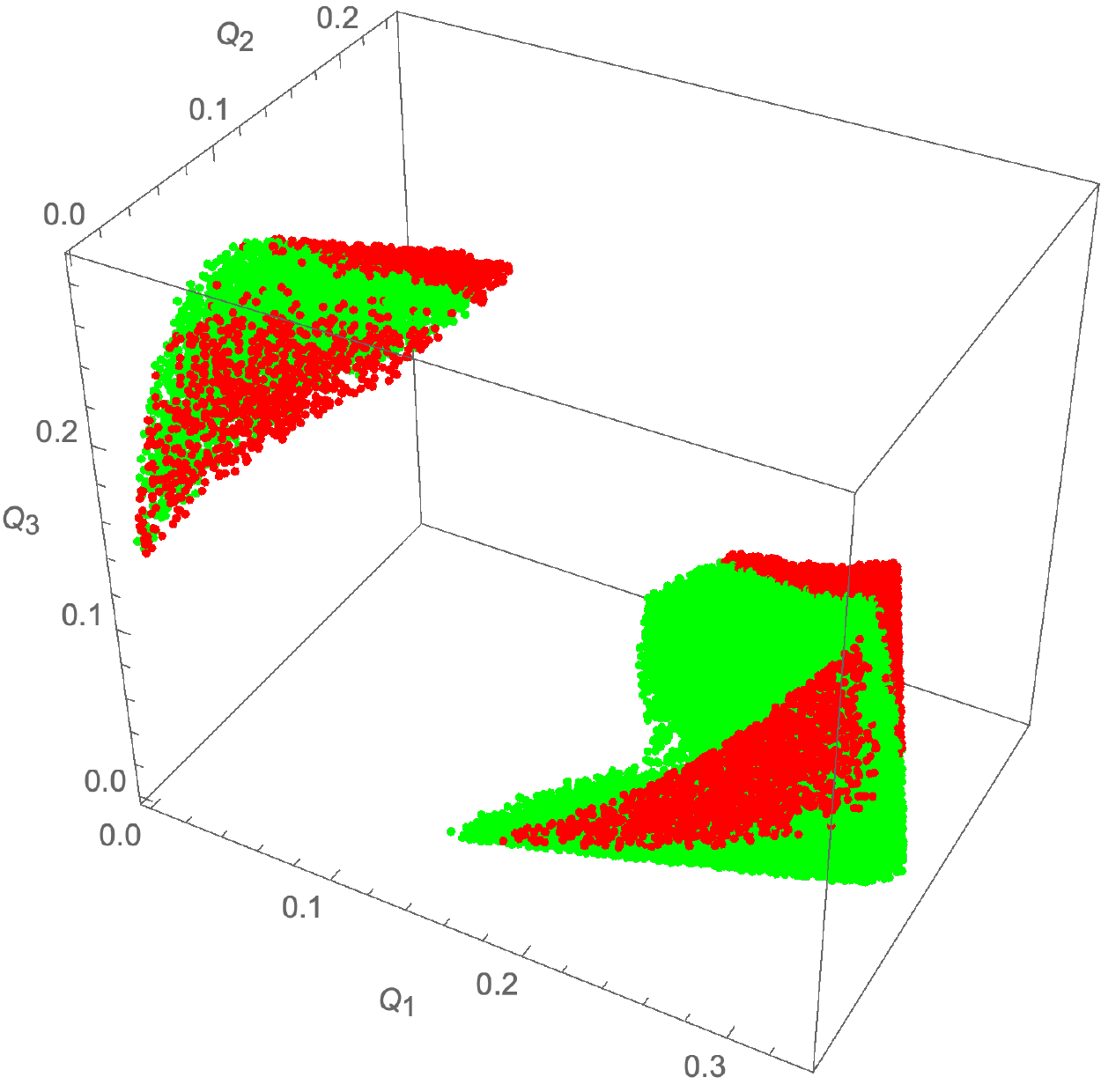}
    \caption{A sampling of just those 
entangled Hiesmayr--L{\"o}ffler  two-qutrit states that {\it do} satisfy the $p>\frac{2^{27}}{3^{18} \cdot 7^{15} \cdot13}$ constraint, but do {\it not} satisfy the $s >\frac{16}{9}$ constraint. The bound-entangled states correspond to the green points, and the free-entangled states to the red. There appear to be {\it two} islands of entanglement.}
    \label{fig:PnotS}
\end{figure}

Now, in a series of figures, let us attempt to gain insight into the specific relations between the constraints and the geometric structure of entanglement. To begin, in Fig.~\ref{fig:PnotS} we show a sampling of just those 
entangled Hiesmayr--L{\"o}ffler  two-qutrit states that {\it do} satisfy the $P \equiv p>\frac{2^{27}}{3^{18} \cdot 7^{15} \cdot13}$ constraint, but do {\it not} satisfy the $S \equiv s >\frac{16}{9}$ constraint. (The sampling is based on use of the Mathematica FindInstance command to generate points satisfying the basic feasible density matrix constraint (\ref{constraint1}), which points are, then, employed to test further constraints. We so proceed, although we are not aware of any particular measure [Hilbert-Schmidt, Bures, \ldots] underlying this command.) The bound-entangled states correspond to the green points, and the free-entangled states to the red. There appear to be {\it two} islands of entanglement.

\begin{figure}
    \centering
    \includegraphics{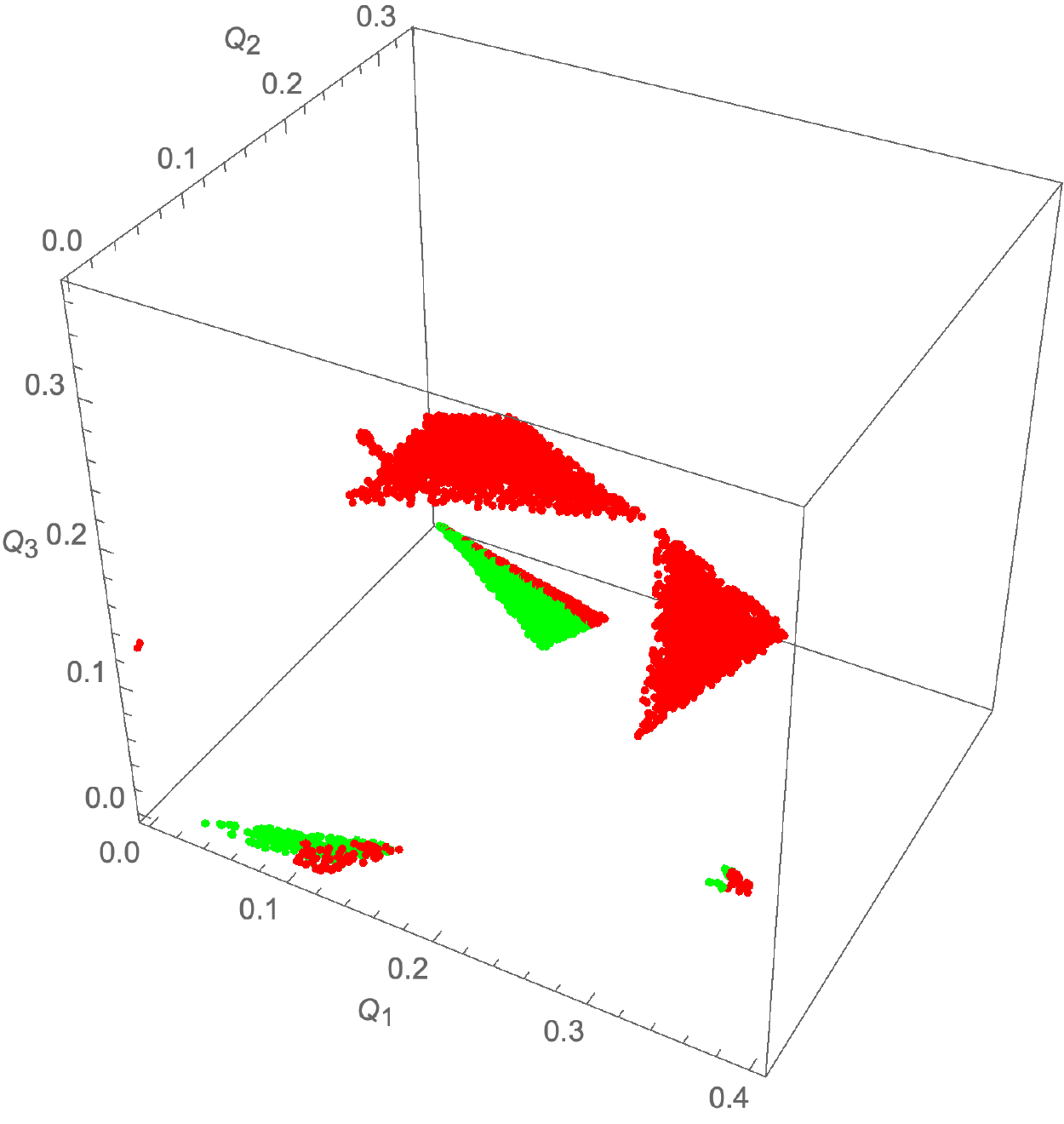}
    \caption{A sampling of just those 
entangled Hiesmayr--L{\"o}ffler  two-qutrit states that do {\it not} satisfy the $p>\frac{2^{27}}{3^{18} \cdot 7^{15} \cdot13}$ constraint, but {\it do} satisfy the $s >\frac{16}{9}$ constraint. The bound-entangled states correspond to the green points, and the free-entangled states to the red. There appear to be multiple islands of entanglement.}
    \label{fig:SnotP}
\end{figure}

In Fig.~\ref{fig:SnotP}, we reverse the role of the two constraints.

\begin{figure}
    \centering
    \includegraphics{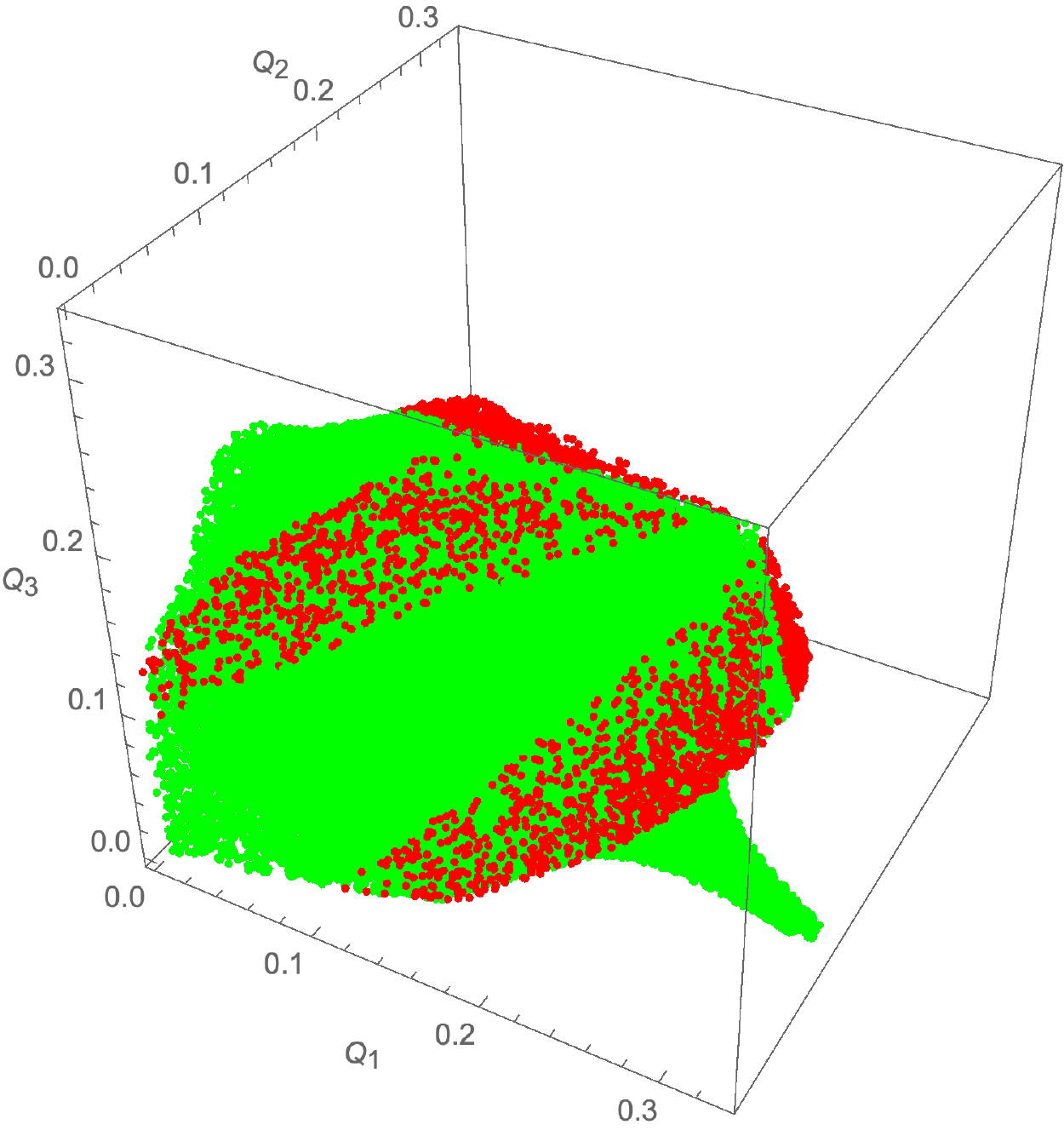}
    \caption{A sampling of those Hiesmayr--L{\"o}ffler  two-qutrit states which satisfy {\it neither} of the entanglement constraints ($P,S$). The green points are separable in nature, while the red ones are  pseudo-one-copy undistillable (POCU) negative partial transposed states. Numerical analyses indicated that for these POCU states, an upper bound on the lowest value that $s$ can attain is 0.47742.}
    \label{fig:POCU}
\end{figure}

Now, in Fig.~\ref{fig:POCU}, we present a sampling of those states which satisfy {\it neither} of the entanglement constraints. The (predominantly) green points are separable in nature, while the red ones appear to be {\it pseudo-one-copy undistillable (POCU) negative partial transposed states} \cite{gabdulin2019investigating}. (``Our results are disclosing that for the two-qutrit system the BE [bound-entangled] states have negligible volume and that these form tiny ‘islands’ sporadically distributed over the surface of the polytope of separable states. The detected families of BE states are found to be located under a layer of pseudo-one-copy undistillable negative partial transposed states with the latter covering the vast majority of the surface of the separable polytope'' \cite{gabdulin2019investigating}. The term ``pseudo'' is used to emphasize that although a single copy of the state is undistillable, a collection of more than one might be.) A Mathematica program is available for testing for the POCU property 
\cite{MandilaraSite}. (One instance of such a point to be so tested is $Q_1=\frac{201}{634},Q_2=\frac{1}{148},
Q_3=\frac{69}{305}$, while another is $Q_1=\frac{761}{2702},Q_2=\frac{3}{422},
Q_3=\frac{47}{290}$.) In fact, employing the indicated  program on a sample of ten candidate POCU states, we were able to confirm that they all possess this property. (Also, all ten $9 \times 9$ density matrices were of full rank.)
Our estimate--using the quasirandom procedure of Martin Roberts \cite{Roberts,Roberts32D,Extreme}--of the Hilbert-Schmidt probability that a Hiesmayr--L{\"o}ffler  two-qutrit state has this 
POCU property is 0.021342868. (Our  calculations in sec.~\ref{Graphic} show that the exact formula for this quantity is 
$\frac{87236}{1061775}+\frac{4 \pi }{27 \sqrt{3}}-\frac{\sqrt{3} \log (2)}{\log
   (81)}-\frac{\cosh ^{-1}(97)}{54 \sqrt{3}} \approx 0.021349$.)

\begin{figure}
    \centering
    \includegraphics{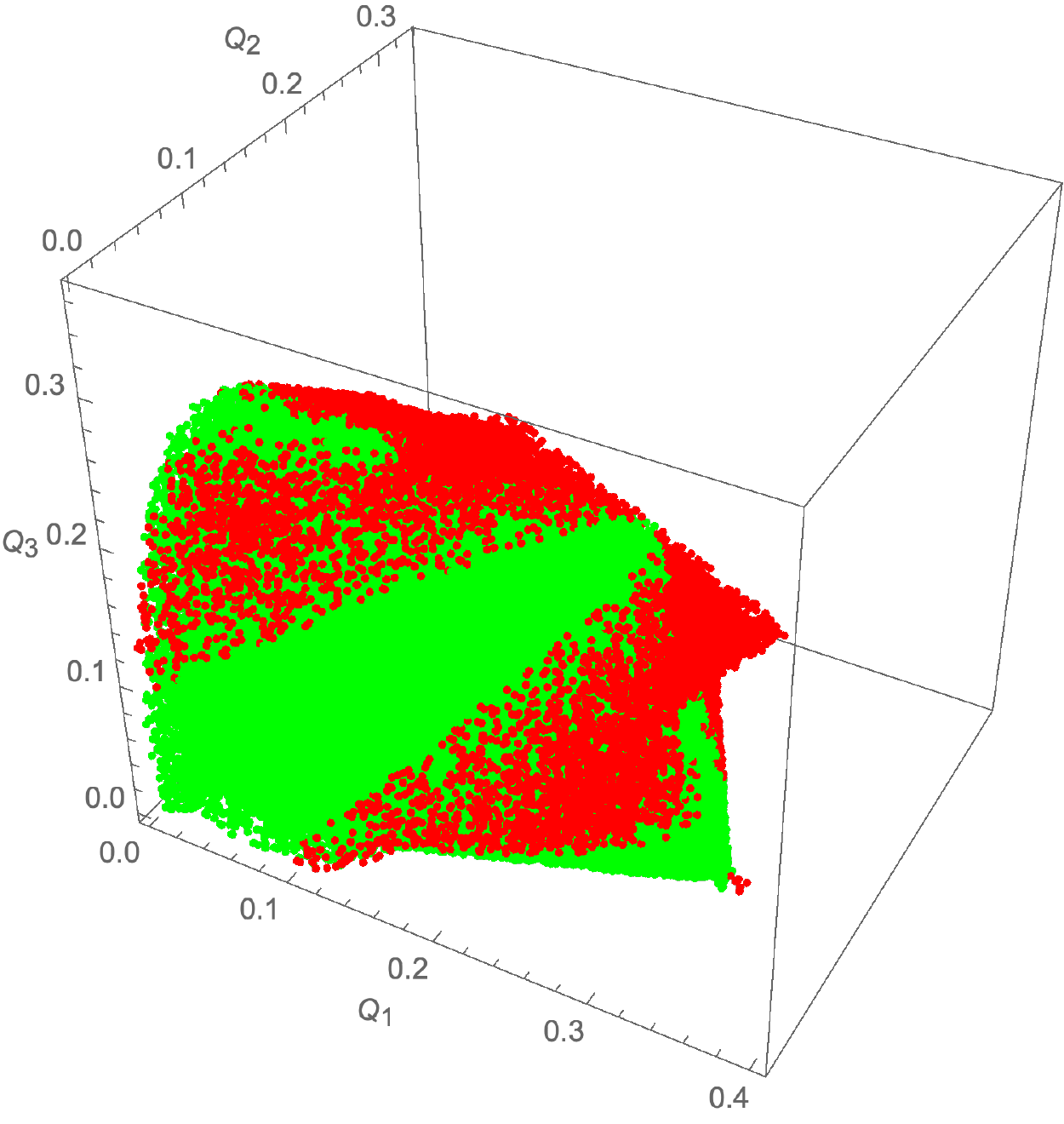}
    \caption{A sampling of those Hiesmayr--L{\"o}ffler  two-qutrit states which do not satisfy at least one of the entanglement constraints ($P,S$). The green points are separable in nature, while the red ones appear to be {\it pseudo-one-copy undistillable negative partial transposed states}. The highest value of $s$ for the red points in this plot is 3.11447.}
    \label{fig:POCU2}
\end{figure}

\begin{figure}
    \centering
    \includegraphics{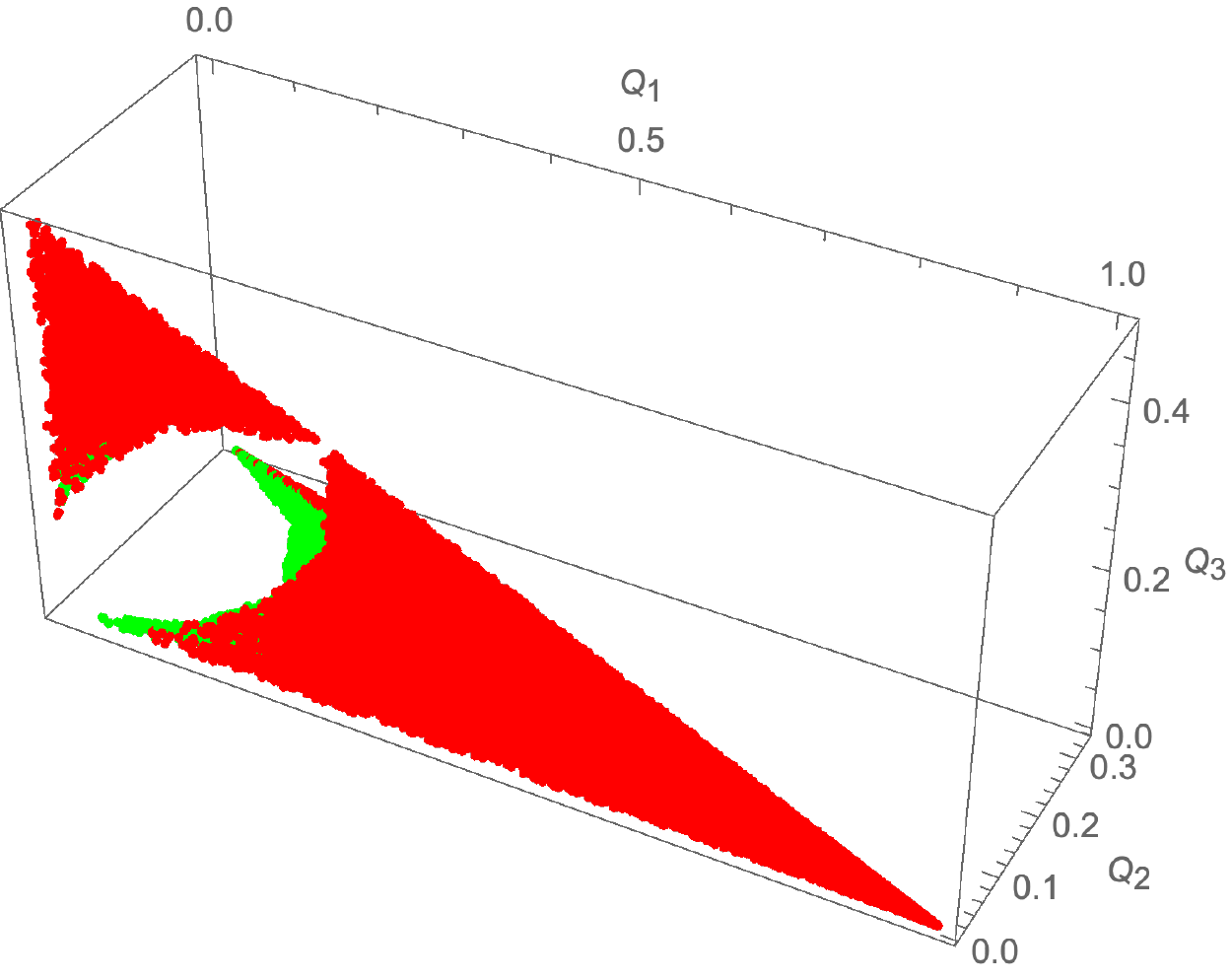}
    \caption{A sampling of those Hiesmayr--L{\"o}ffler  two-qutrit states which  satisfy at least one of the entanglement constraints. The bound-entangled states correspond to the green points, and the free-entangled states to the red.}
    \label{fig:PorS}
\end{figure}

\begin{figure}
    \centering
    \includegraphics{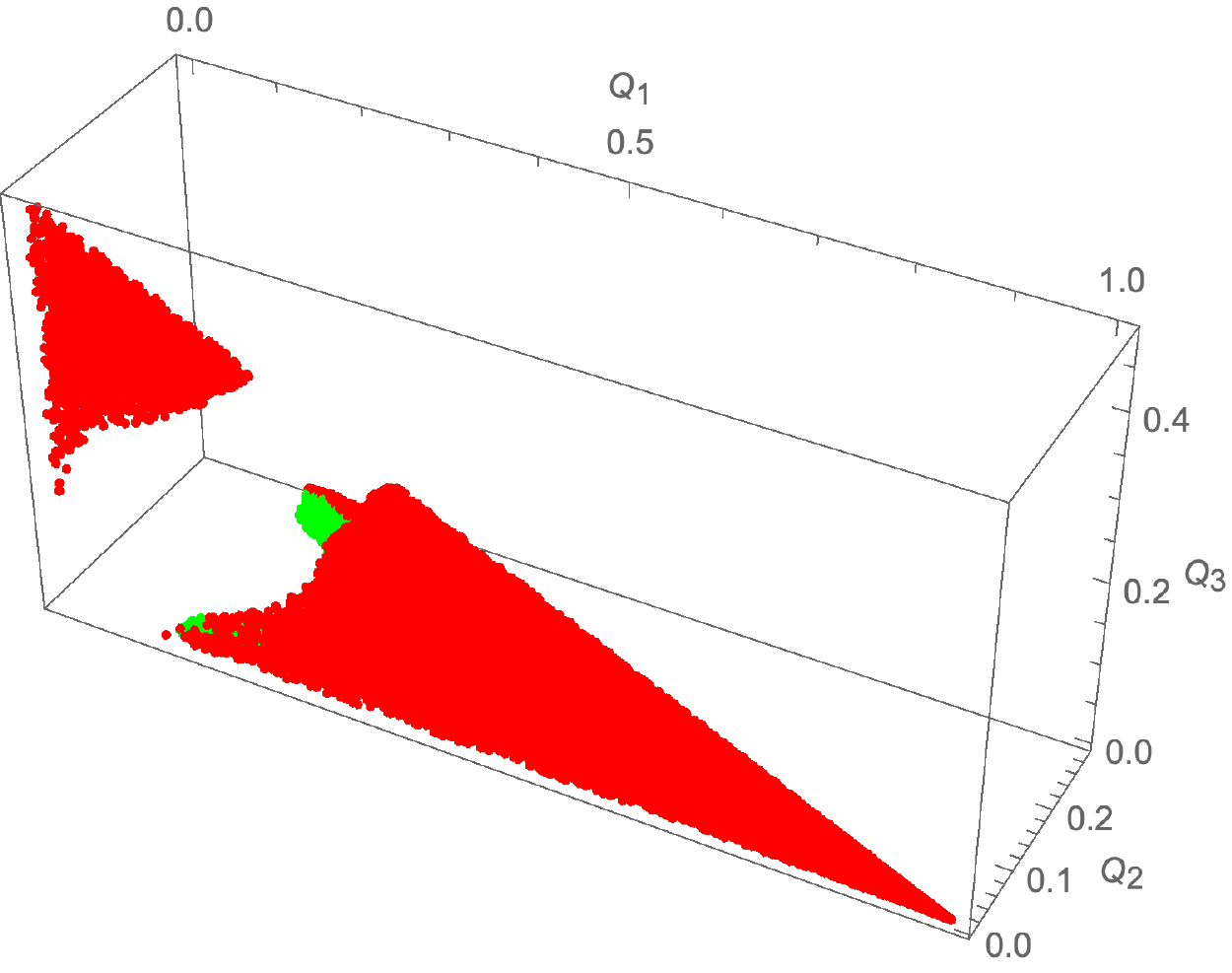}
    \caption{A sampling of those states which satisfy {\it both} of the entanglement constraints. The bound-entangled states correspond to the green points, and the free-entangled states to the red.}
    \label{fig:PandS}
\end{figure}

Numerical analyses indicated that for these POCU states, an upper bound on the lowest value that $s$ can attain is 0.47742 (at $Q_1=\frac{16022}{89351}, Q_2=\frac{28}{185}, Q_3=\frac{101}{551}$).
In Figs.~\ref{fig:POCU2}, \ref{fig:PorS} and \ref{fig:PandS}, we show plots based on additional Boolean combinations of the two constraints. (Note that there are some differences in scaling among the several figures in the paper.)
\subsubsection{States on the boundary of separability}
\begin{figure}
    \centering
    \includegraphics{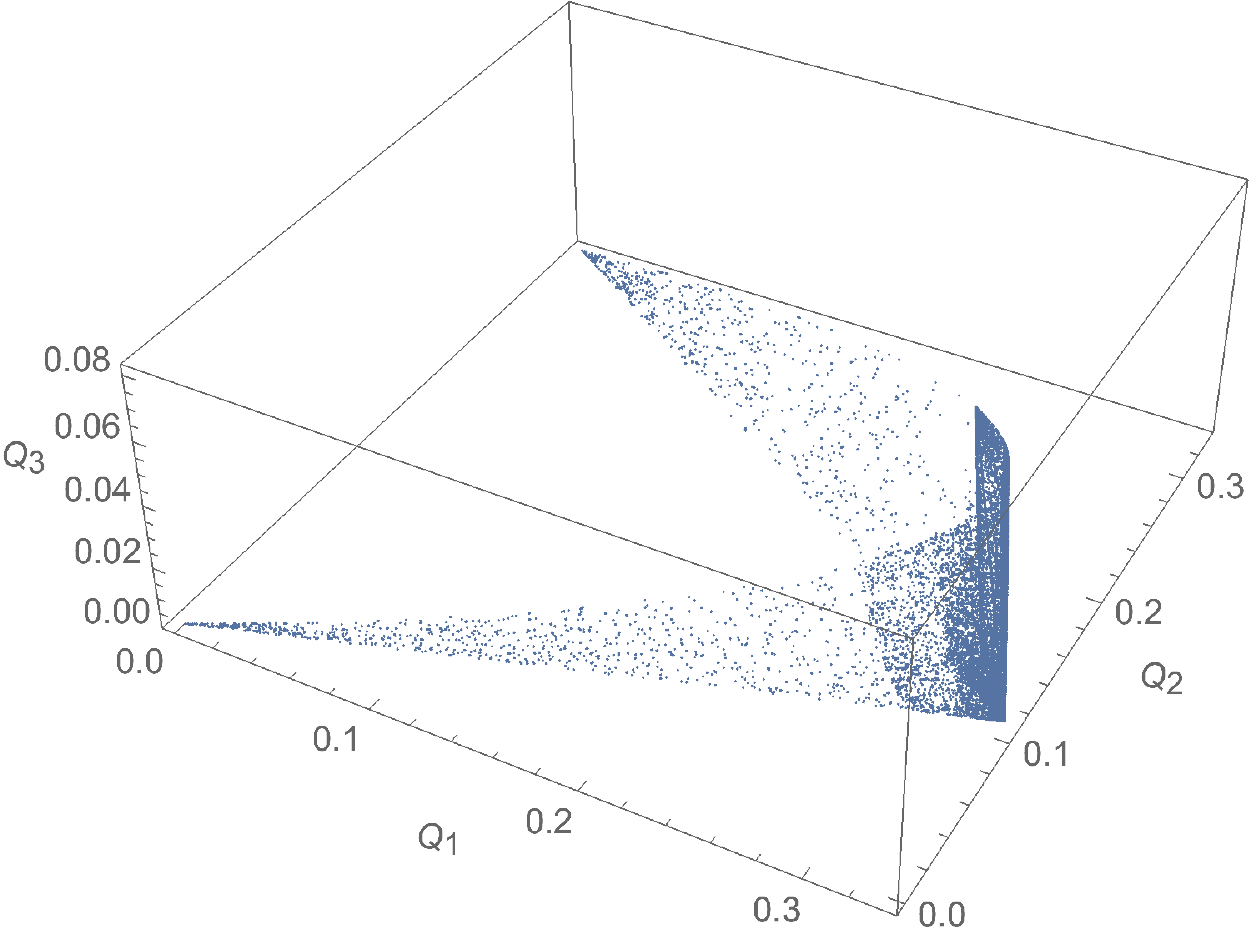}
    \caption{Hiesmayr--L{\"o}ffler  two-qutrit states on the boundary of the separable states for which the $S$ entanglement constraint is saturated, i. e. $s=\frac{16}{9}$.}
    \label{fig:PPTnoboundarySsaturated}
\end{figure}

\begin{figure}
    \centering
    \includegraphics{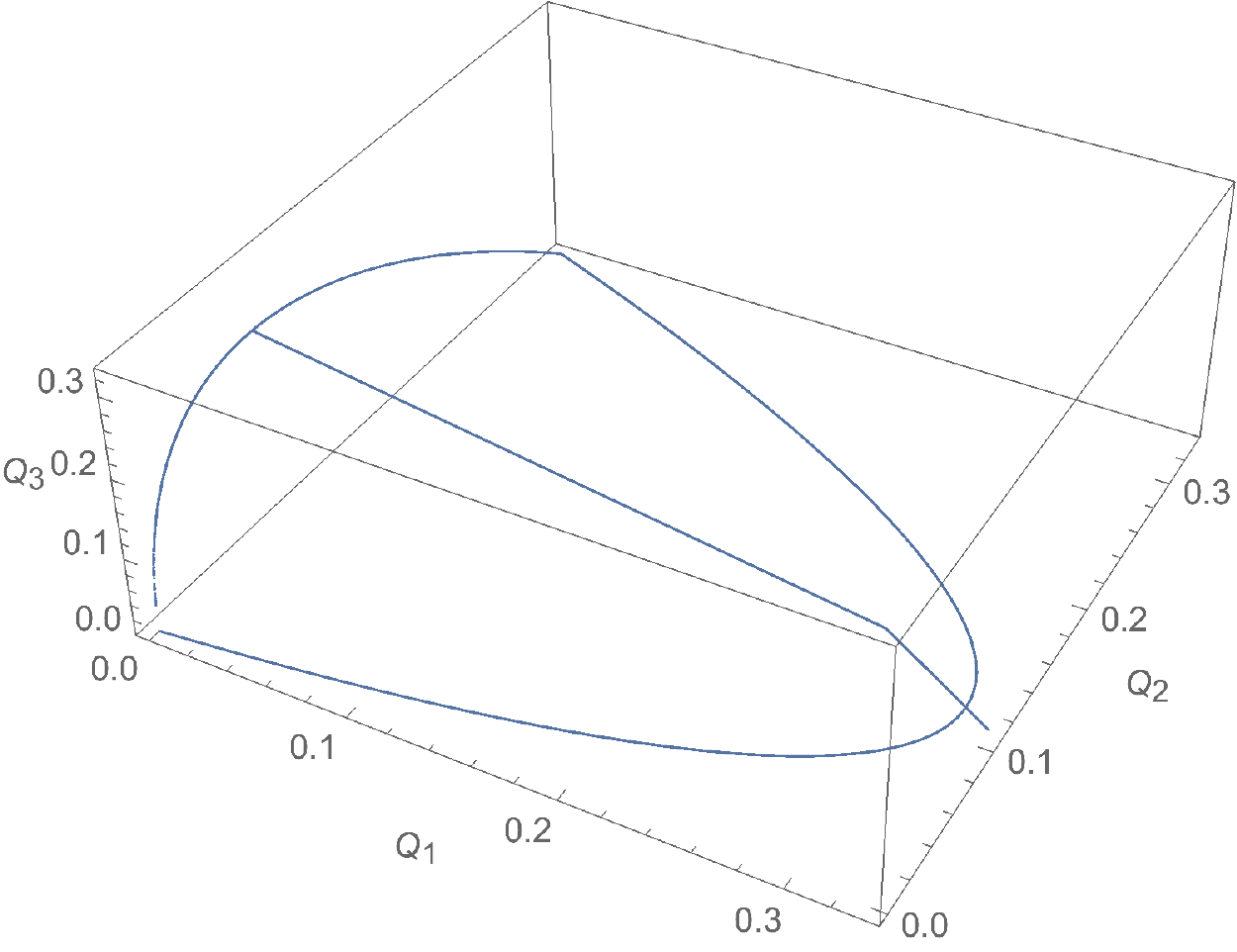}
    \caption{Hiesmayr--L{\"o}ffler  two-qutrit states on the boundaries of both the separable states and PPT states for which the $S$ entanglement constraint is saturated, i. e. $s=\frac{16}{9}$.}
    \label{fig:PPTboundarySsatisfied}
\end{figure}

\begin{figure}
    \centering
    \includegraphics{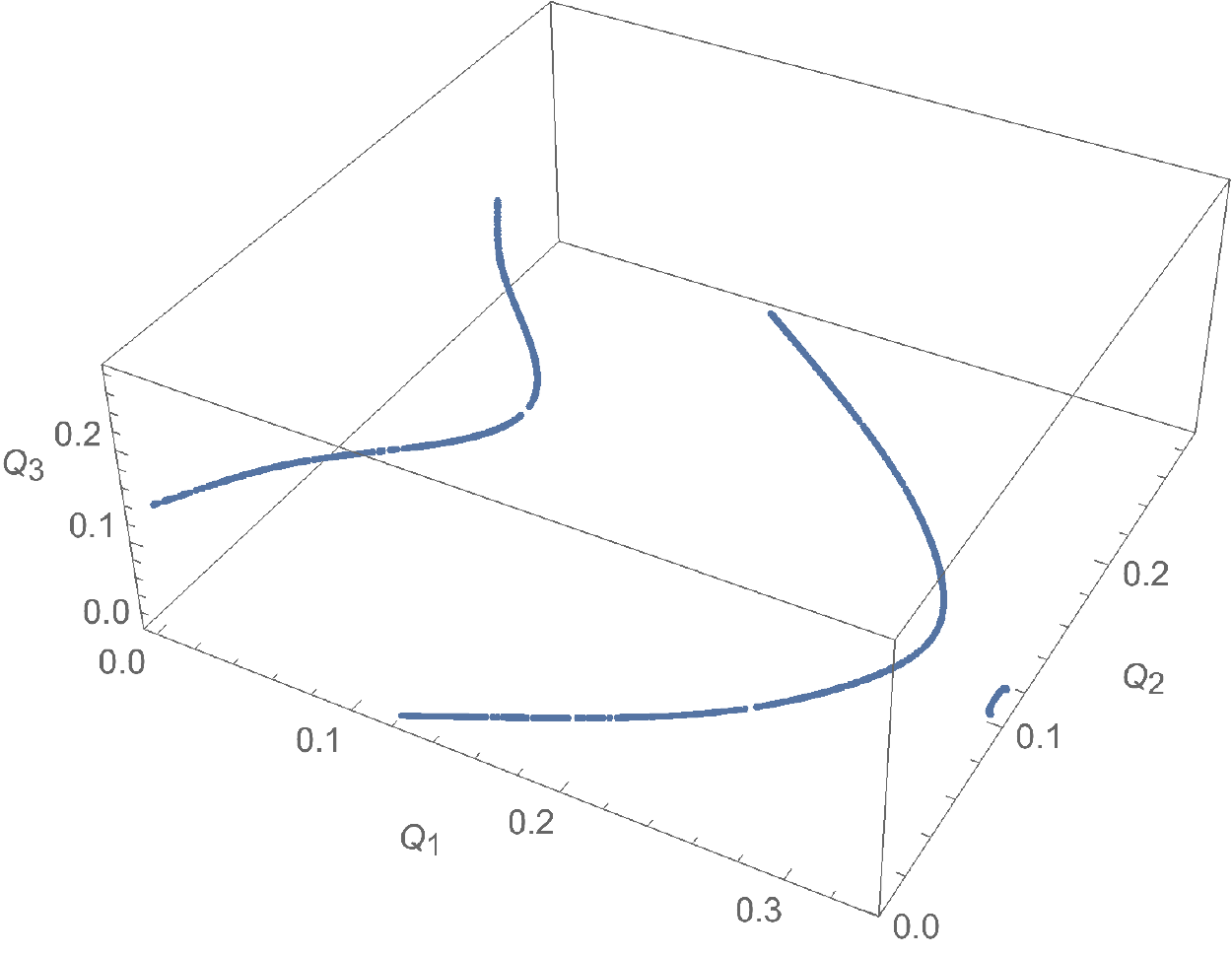}
    \caption{Hiesmayr--L{\"o}ffler  two-qutrit states on the boundaries of both the separable states and PPT states for which the $P$ entanglement constraint is saturated, i. e. $p=\frac{134217728}{23910933822616040487651}=\frac{2^{27}}{3^{18} \cdot 7^{15} \cdot13} \approx 5.61324 \cdot 10^{-15}$. There are three curves, one much smaller than the other two.}
    \label{fig:PPTboundaryPsatisfied}
\end{figure}

\begin{figure}
    \centering
    \includegraphics{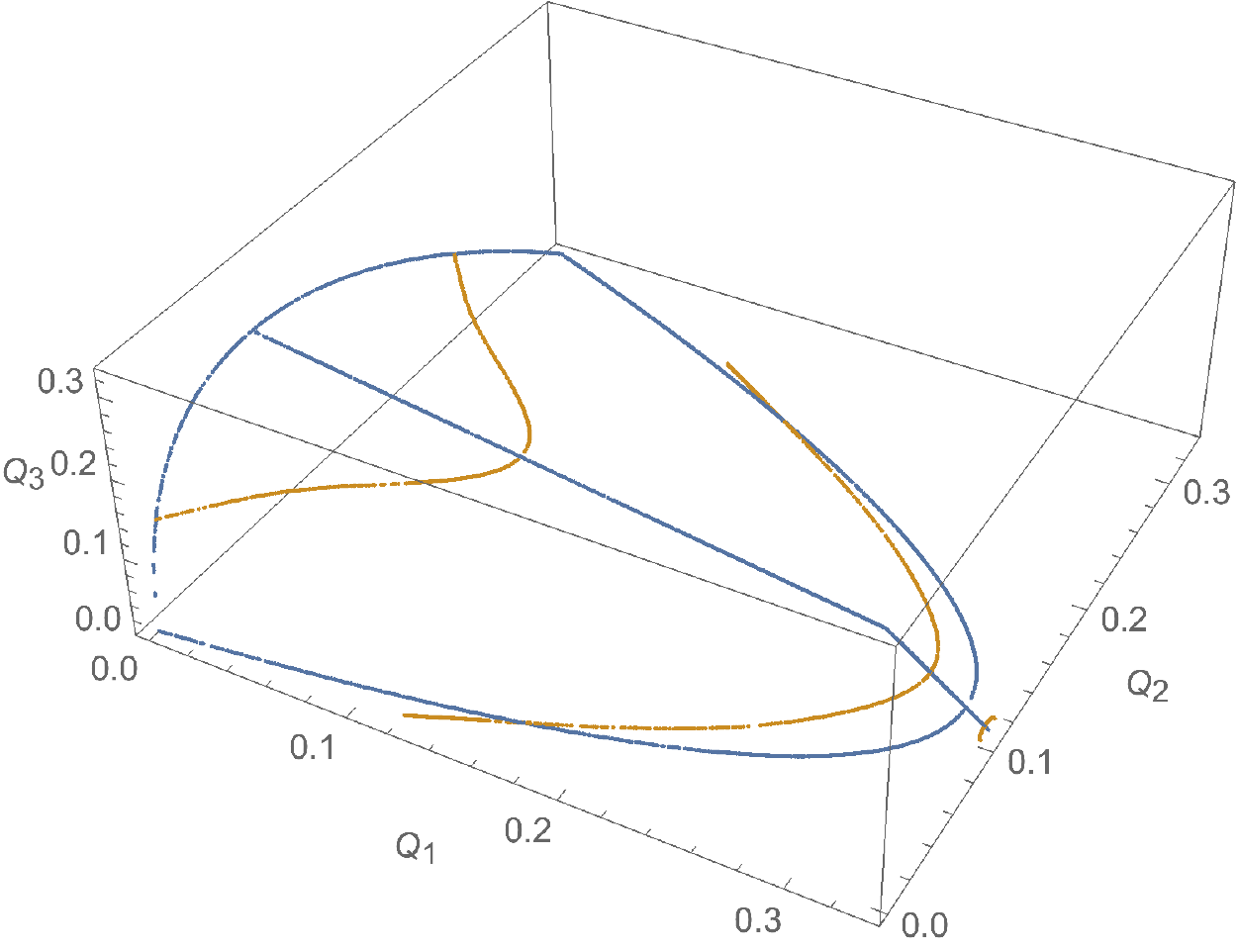}
    \caption{A joint plot of the two curves (Fig.~\ref{fig:PPTboundarySsatisfied} and Fig.~\ref{fig:PPTboundaryPsatisfied}), showing the intersection of the PPT boundary of the Hiesmayr--L{\"o}ffler  two-qutrit states with points for which $s=\frac{16}{9}$ and $p=\frac{2^{27}}{3^{18} \cdot 7^{15} \cdot13}$, respectively.}
    \label{fig:Combined}
\end{figure}

The points in the next two figures (Figs.~\ref{fig:PPTnoboundarySsaturated} and \ref{fig:PPTboundarySsatisfied}) all saturate the $S$ entanglement constraint, i. e., $s=\frac{16}{9}$. The points in the former lie, in general, within the PPT states, while in the latter, they lie on the boundary of the PPT states.
Efforts of ours to produce a companion pair of figures to these last two, in which instead of the $S$ entanglement constraint being saturated, the $P$ constraint would be, proved much more computationally challenging. However, we were able to obtain a fewer-point analogue of Fig.~\ref{fig:PPTboundarySsatisfied}, that is, Fig.~\ref{fig:PPTboundaryPsatisfied}.
In Fig.~\ref{fig:Combined} we jointly plot the two curves (Fig.~\ref{fig:PPTboundarySsatisfied} and Fig.~\ref{fig:PPTboundaryPsatisfied}), showing the intersection of the PPT boundary with points saturating the $S$ and $P$ constraints, respectively.

In Table~\ref{tab:Second}, we summarize several of our analyses.
\begin{table}
\caption{\label{tab:Second}Exact formulas and underlying quasirandom estimates 
\cite{Roberts,Roberts32D,Extreme,slater2019pair} of various Hilbert-Schmidt probabilities for the Hiesmayr-L{\"o}ffler $d=3$ two-qutrit model. More than four hundred million points were employed--and results appear accurate to six-seven decimal places. $S$ denotes the set satisfying the constraint $s> \frac{16}{9} \approx 1.7777$ and $P$, the constraint $p>\frac{134217728}{23910933822616040487651}=\frac{2^{27}}{3^{18} \cdot 7^{15} \cdot13} \approx 5.61324 \cdot 10^{-15}$.  Notationally, $\neg$ is the negation logic operator (NOT); $\land$ is the conjunction logic operator (AND); and $\lor$  is the disjunction logic operator (OR). Alternative integration procedures to quasirandom estimation were used for the last two entries.  (Somewhat interesting observations with regard to the entries of the revised table are that $\cosh ^{-1}(97)= \log \left(97+56 \sqrt{3}\right)=\sinh ^{-1}\left(56 \sqrt{3}\right)$, so that $\sqrt{3}$ is even more omnipresent.)}
$\left(
\begin{array}{ccc}
\hline
 Set & Probability  & Quasirandom Estimate  \\ 
 \hline
 \hline
\_ & 1 & 1.0000000 \\
 \text{PPT} & \frac{8 \pi }{27 \sqrt{3}} & 0.53742158 \\
 \neg P\land \neg S & \frac{21}{44} & 0.47726800 \\
 P & \frac{4702531}{4247100}-\frac{4 \pi }{27 \sqrt{3}}-\frac{\sqrt{3} \log (2)}{\log
   (81)}-\frac{\cosh ^{-1}(97)}{54 \sqrt{3}} & 0.50900327 \\
 S & \frac{1}{81} \left(27+\sqrt{3} \log \left(97+56 \sqrt{3}\right)\right) & 0.44597788
   \\
 P\land S & \frac{974539}{1061775}-\frac{4 \pi }{27 \sqrt{3}}-\frac{\sqrt{3} \log (2)}{\log
   (81)}+\frac{\cosh ^{-1}(97)}{54 \sqrt{3}} & 0.43224916 \\
 P\lor S & \frac{23}{44} & 0.52273200 \\
 \neg P\lor \neg S &   \frac{1678081}{4247100}-\frac{4 \pi }{27 \sqrt{3}}+\frac{\sqrt{3} \log (2)}{\log
   (81)}+\frac{\cosh ^{-1}(97)}{54 \sqrt{3}} & 0.56775084 \\
 \text{PPT}\land \neg P\land \neg S & \frac{1678081}{4247100}-\frac{4 \pi }{27 \sqrt{3}}+\frac{\sqrt{3} \log (2)}{\log
   (81)}+\frac{\cosh ^{-1}(97)}{54 \sqrt{3}} & 0.45591798 \\
\text{PPT}\land P & \frac{54029}{386100}+\frac{4 \pi }{27 \sqrt{3}}-\frac{\sqrt{3} \log (2)}{\log
   (81)}-\frac{\cosh ^{-1}(97)}{54 \sqrt{3}} & 0.079128512 \\
 \text{PPT}\land S & \frac{2}{81} \left(4 \sqrt{3} \pi -21\right) & 0.018903658 \\
 \text{PPT}\land P\land S & \frac{2}{121} & 0.016528575 \\
 \text{PPT}\land (P\lor S) & -\frac{1678081}{4247100}+\frac{4 \pi }{9 \sqrt{3}}-\frac{\sqrt{3} \log (2)}{\log
   (81)}-\frac{\cosh ^{-1}(97)}{54 \sqrt{3}} & 0.081503595 \\
 \text{PPT}\land (\neg P\lor \neg S) &   \frac{8 \pi }{27 \sqrt{3}}-\frac{2}{121} & 0.52089300 \\
 \hline
 \text{PPT}\land S\land \neg P & \frac{4 \left(242 \sqrt{3} \pi -1311\right)}{9801} & 0.002374589709\\
 \neg \text{PPT}\lor S & \frac{13}{27} & 0.48148148 \\
\end{array}
\right)$
\end{table}
\subsection{Boolean-analysis-based derivation of the formulas in Table~\ref{tab:Second}}
The formulas in this table were derived making use of the decomposition into eight ``atoms'' of the 256-dimensional algebra associated with the three sets $PPT, P,S$. We now present the final answer to \cite{user250938}--omitting the already-presented Table~\ref{tab:Second}--discussing the underlying analysis (in terms of the notation in \cite{user250938}, $A \equiv P, B \equiv P, C \equiv PPT$):

``We determine--making strong use of the Mathematica code given by user250938 in the answer to this question--the eight atoms of our 256-dimensional Boolean algebra on three sets. Then, we are  able to present a  table of imposed constraints and their (now partially revised) associated probabilities fully consistent with this framework.

(The several integer denominators [in Table~\ref{tab:Second}] all have prime factorizations with primes no greater than 13--but certainly not the numerators. The prime 97 plays a conspicuous role.)

To obtain these results, we began by estimating the values of the eight atoms--in the indicated order 
\begin{equation} \label{8atoms}
P \land S \land PPT, \neg P \land S \land PPT, P \land \neg S \land PPT,P \land S \land \neg PPT, \neg P \land \neg S \land PPT,
\end{equation}
\begin{displaymath}
\neg P \land S \land \neg PPT,P \land \neg S \land \neg PPT,\neg P \land \neg S \land \neg PPT
\end{displaymath}
as--
\begin{equation}
 \{\frac{2984353}{180555569},\frac{428757}{180555569},\frac{11302706}{180555569},\frac{75060766}{180555569},\frac{82318620}{180555569},\frac{2050053}{180555569},\frac{2555632}{180555569},\frac{3854682}{180555569}\}   
\end{equation}
\begin{displaymath}
\approx \{0.01652872308,0.002374653977,0.06259959780,0.4157211346,0.4559184768,
\end{displaymath}
\begin{displaymath}
0.01135413885,0.0
   1415426848,0.02134900641\}.
\end{displaymath}

The estimation procedure employed was the "quasirandom" ("generalized golden ratio") one of Martin Roberts https://math.stackexchange.com/questions/2231391/how-can-one-generate-an-open-ended-sequence-of-low-discrepancy-points-in-3d .
It was used to generate six-and-a half billion points (triplets in $[0,1]^3$), only approximately one-thirty-sixth of them--those yielding feasible density matrices--being further utilized.

These eight estimated values (summing to 1) are well fitted, we find (using the Mathematica Solve command), by

$\frac{2}{121},\frac{4 \left(242 \sqrt{3} \pi
   -1311\right)}{9801},\frac{524119}{4247100}+\frac{4 \pi }{27 \sqrt{3}}-\frac{\sqrt{3}
   \log (2)}{\log (81)}-\frac{\cosh ^{-1}(97)}{54 \sqrt{3}}
,\frac{7909}{8775}-\frac{4
   \pi }{27 \sqrt{3}}-\frac{\sqrt{3} \log (2)}{\log (81)}+\frac{\cosh ^{-1}(97)}{54
   \sqrt{3}},\frac{1678081}{4247100}-\frac{4 \pi }{27 \sqrt{3}}+\frac{\sqrt{3} \log
   (2)}{\log (81)}+\frac{\cosh ^{-1}(97)}{54 \sqrt{3}},$
$-\frac{434}{8775}-\frac{4 \pi
   }{27 \sqrt{3}}+\frac{\sqrt{3} \log (2)}{\log (81)}+\frac{\cosh ^{-1}(97)}{54
   \sqrt{3}},\frac{70064}{1061775}-\frac{4 \pi }{27 \sqrt{3}}+\frac{\sqrt{3} \log
   (2)}{\log (81)}-\frac{\cosh ^{-1}(97)}{54 \sqrt{3}},\frac{87236}{1061775}+\frac{4 \pi
   }{27 \sqrt{3}}-\frac{\sqrt{3} \log (2)}{\log (81)}-\frac{\cosh ^{-1}(97)}{54
   \sqrt{3}}$.
   $\approx 0.01652892562,0.002374589709,0.06259481829,0.4157208527,0.4559237002$,
   
   $0.01135281657,0.0
   1415526980,0.02134902704$.

To get these formulas yielded by Solve  for the eight atoms, we first incorporated into the analysis, the three results--$\frac{8 \pi }{27 \sqrt{3}},\frac{1}{81} \left(27+\sqrt{3} \log \left(97+56
   \sqrt{3}\right)\right),\frac{2}{81} \left(4 \sqrt{3} \pi -21\right)$--having earlier  been obtained \cite{slater2019bound} through symbolic integration. Then, having strong confidence in the previously (tabulated)  used values of $\frac{21}{44},\frac{2}{121}$ and $\frac{8 \pi }{27 \sqrt{3}}-\frac{2}{121}$ expressions, we incorporated them too. 

Since these six values were not fully sufficient for Solve, we additionally employed the WolframAlpha site--searching over the 256 BooleanFunction results to find simple well-fitting formulas, using the above-given numerically estimated values of the eight atoms. For instance, for BooleanFunction[133,{P,S,PPT}]=$(P \land PPT \land S) \lor (\neg P \land \neg PPT)$, the site suggested $\frac{16}{325}$, fitting the estimated corresponding value to a ratio of 1.00000006615. Also, for  BooleanFunction[62,{P,S,PPT}]=$\neg (P \land S) \land (P \lor PPT \lor S)$, the suggestion was $\frac{\sqrt{3} \log (2)}{\log (9)}$, having an analogous ratio of 0.999999807781. 

Incorporating as well, these last two results, as well as the previously tabulated $\frac{13}{27}$ for $\neg PPT \lor S$, proved sufficient to obtain the eight ``atomic" formulas.

The close-to-1 ratios of these formulas to the estimated values, given above, are 
$\{1.000012254,0.9999729358,0.9999236495,0.9999993220$,
$1.000011457,0.9998835421,1.000070743,1.000000966\}$ ."

\begin{figure}
    \centering
    \includegraphics{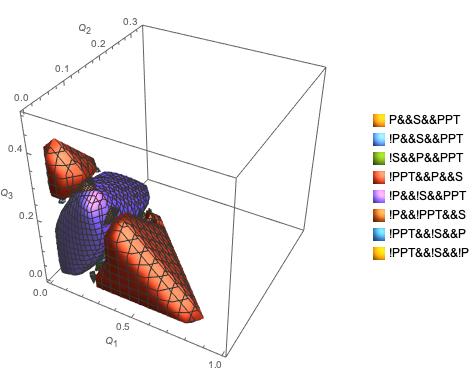}
    \caption{Decomposition of the Hiesmayr--L{\"o}ffler  two-qutrit states into its mutually exclusive eight atoms. The PPT states are in the interior of the body and the entangled states at the extremities. Exclamation signs in the legend denote set negation, and the double ampersand, set intersection.}
    \label{fig:Venn}
\end{figure}

In Fig.~\ref{fig:Venn}, we now, additionally, display the eight atoms spanning the entanglement-probability boolean algebra for the Hiesmayr--L{\"o}ffler  two-qutrit model.
(We have also investigated--as a supplement to Fig.~\ref{fig:Venn}--the potential use of [planar] Venn diagrams to represent the various entanglement-related probabilities associated with the boolean combinations of $P,S$ and $PPT$ \cite{kgir}.)

\subsection{Analyses employing Li-Qiao variables $\alpha_i,\beta_i$}
As a matter of analytical interest, we had initially 
concentrated  upon attempting to construct the proper entanglement bounds--now reported above--for $P$ and $S$ applicable to the Hiesmayr--L{\"o}ffler  two-qutrit model, but strictly within the Li-Qiao framework.
In so doing, we follow  \cite{slater2020archipelagos2}, in which we employed the  well-known necessary and sufficient conditions for nonnegative-semidefiniteness that all leading minors  be nonnegative \cite{prussing1986principal}. There are twenty-two sets of such minors of $3 \times 3$ density matrices to so consider, since the Li-Qiao algorithm expands $\rho_{HL}$ into eleven separable two-qutrit states. (We were able to obtain this explicit expansion, lending us confidence in our further analyses.
In the Li-Qiao setup, we initially have twenty parameters, ten $\alpha_i$ and ten $\beta_i$, with $t_i= \alpha_i \beta_i$.
Then, the solution yielding the correct expansion was expressible as $\beta_i=\frac{2 (Q_1-Q_3)}{3 \alpha_i}, i=1,4,6$ and 
$\beta_i=-\frac{2 (Q_1-Q_3)}{3 \alpha_i}, i=2,5,7$, and $\beta_i=\frac{-1+3 Q_1+6 Q_3}{3 \alpha_i}, i=3,8$, and 
$\beta_i=\frac{-1+Q_1+6 Q_2 +2 Q_3}{3 \alpha_i}, i=9, 10$.)

Then, using numerical integration in a {\it thirteen}-dimensional setting ($Q_1, Q_2, Q_3$ and the ten  $\alpha_i$'s), our highest estimate of the (multiplicative) bound for $P$ was $p=8.91229*10^{-22}$, and of the (additive) bound for $S$ was $s=0.155322$. 
Requiring that $p>8.91229*10^{-22}$, yields an entanglement probability estimate of 0.764984, and enforcing $s>0.155322$, gives 0.972243. So, these bounds are disappointingly small, leading to entanglement probability estimates clearly too large, given the known PPT probability $\frac{8 \pi }{27 \sqrt{3}} \approx 0.537422$, all but only $\frac{2}{81} \left(4 \sqrt{3} \pi -21\right) \approx 0.0189305$  of which is bound-entangled, through enforcement of the realignment test.

So, while we are confidently able to claim knowledge of the proper bounds for the pair of Li-Qiao entanglement constraints on the singular-value-based terms $s$ and $p$ for the Hiesmayr-L{\"o}ffler  two-qutrit magic simplex of Bell states, this was only achievable in the first of our two lines of two-qutrit analysis, employing simply the trivariate ($Q_1, Q_2, Q_3$) set of constraints ((\ref{constraint1})-(\ref{realignment})). The second line of  13-variable ($Q_1, Q_2, Q_3$ and ten Li-Qiao parameters $\alpha_i, i=1,\cdots,10$) analyses, conducted within the Li-Qiao framework, had not similarly  succeeded.
\subsubsection{Explicit decompositions of separable states}
However, further analyses allowed us to construct multiple (about twenty, presently) sets of Li-Qiao parameters $\alpha_i, \beta_i, i =1,\ldots,10$ (cf. \cite[eqs. (63)-(66)]{li2018separable}) each yielding a separable expansion of length eleven (each component product density matrix  being equally weighted by $\frac{1}{11}$) for specific Hiesmayr-L{\"o}ffler  two-qutrit states. For example, the ten $\alpha$ parameters $\left\{\frac{25}{256},\frac{75}{512},\frac{35}{256},\frac{171}{10},\frac{15}{128},-\frac{
   103}{5},-\frac{5}{256},\frac{55}{512},\frac{25}{512},-\frac{15}{512}\right\}$, together with the ten $\beta$ parameters
   $\left\{-\frac{15}{2},-\frac{31}{2},\frac{55}{512},-\frac{5}{64},\frac{4}{5},-\frac{5}{256
   },\frac{187}{10},\frac{29}{5},0,-\frac{101}{10}\right\}$ gave us a separable decomposition for the state with 
   $Q_1,=\frac{136847}{1179648},Q_2=\frac{256369}{2359296},Q_3=\frac{136847}{1179648}$. Somewhat disappointingly however, all the twenty-or-so examples so far generated had $Q_1=Q_3$, so the multiplicative norm (\ref{MultiplicativeNorm}) simply reduced to zero. The greatest value for the additive norm (\ref{AdditiveNorm}) so far generated is $\frac{18225}{16777216} \approx 0.00108629 <\frac{16}{9} \approx 1.7777$.
\subsection{Best separable approximation}
In their pair of recent skillful papers \cite{li2018necessary,li2018separable}, Li and Qiao presented necessary {\it and} sufficient conditions for separability, the implementation of which we have investigated above. They did not, however, discuss the apparently related {\it best separable approximation} problem \cite{akulin2015essentially}. To begin a study of the possible  application of the Li-Qiao analytical framework to this problem of major interest, we sought a best separable approximation for the entangled  Hiesmayr-L{\"o}ffler two-qutrit density matrix (\ref{d=3HL}) with its parameters having been set to $Q_1=\frac{4235}{50001},Q_2=\frac{1}{166},
Q_3=\frac{30}{113}$. Then, we obtained a value of 
$B=0.195662$,  where $B$ is the parameter one seeks to minimize $0 \leq B \leq 1$, in the equation 
\cite[eq. (2)]{gabdulin2019investigating}
\begin{equation} \label{BSA}
\hat{\rho}   =(1-B){\hat{\rho}_{sep}}  +B {\hat{\rho}_{ent}} .
\end{equation}
Now, the minimum $B=0.195662$ for the indicated choice of $Q$'s for ${\hat{\rho}}$ is obtained if we choose for the parameterization of 
${\hat{\rho}_{ent}}$, the values $Q_1=1.50726 \cdot 10^{-7},Q_2=1.95701 \cdot 10^{-8}$ and $Q_3=0.5$.  Then, from (\ref{BSA}), we can obtain the desired ${\hat{\rho}_{sep}}$--for which $Q_1=0.10754,Q_2=0.0074895$ and $Q_3=0.208439$.
\section{Two-ququart analyses} \label{twoququart}
For the $d=4$ two-ququart 
Hiesmayr-L{\"o}ffler magic simplex states, 
\begin{equation} \label{d=4HL}
\rho_{HL}^{2qq}= \left(
\begin{array}{cccccccccccccccc}
 \kappa _1 & 0 & 0 & 0 & 0 & \kappa _2 & 0 & 0 & 0 & 0 & \kappa _2 & 0 & 0 & 0 & 0 &
   \kappa _2 \\
 0 & Q_2 & 0 & 0 & 0 & 0 & 0 & 0 & 0 & 0 & 0 & 0 & 0 & 0 & 0 & 0 \\
 0 & 0 & Q_3 & 0 & 0 & 0 & 0 & 0 & 0 & 0 & 0 & 0 & 0 & 0 & 0 & 0 \\
 0 & 0 & 0 & \kappa _3 & 0 & 0 & 0 & 0 & 0 & 0 & 0 & 0 & 0 & 0 & 0 & 0 \\
 0 & 0 & 0 & 0 & \kappa _3 & 0 & 0 & 0 & 0 & 0 & 0 & 0 & 0 & 0 & 0 & 0 \\
 \kappa _2 & 0 & 0 & 0 & 0 & \kappa _1 & 0 & 0 & 0 & 0 & \kappa _2 & 0 & 0 & 0 & 0 &
   \kappa _2 \\
 0 & 0 & 0 & 0 & 0 & 0 & Q_2 & 0 & 0 & 0 & 0 & 0 & 0 & 0 & 0 & 0 \\
 0 & 0 & 0 & 0 & 0 & 0 & 0 & Q_3 & 0 & 0 & 0 & 0 & 0 & 0 & 0 & 0 \\
 0 & 0 & 0 & 0 & 0 & 0 & 0 & 0 & Q_3 & 0 & 0 & 0 & 0 & 0 & 0 & 0 \\
 0 & 0 & 0 & 0 & 0 & 0 & 0 & 0 & 0 & \kappa _3 & 0 & 0 & 0 & 0 & 0 & 0 \\
 \kappa _2 & 0 & 0 & 0 & 0 & \kappa _2 & 0 & 0 & 0 & 0 & \kappa _1 & 0 & 0 & 0 & 0 &
   \kappa _2 \\
 0 & 0 & 0 & 0 & 0 & 0 & 0 & 0 & 0 & 0 & 0 & Q_2 & 0 & 0 & 0 & 0 \\
 0 & 0 & 0 & 0 & 0 & 0 & 0 & 0 & 0 & 0 & 0 & 0 & Q_2 & 0 & 0 & 0 \\
 0 & 0 & 0 & 0 & 0 & 0 & 0 & 0 & 0 & 0 & 0 & 0 & 0 & Q_3 & 0 & 0 \\
 0 & 0 & 0 & 0 & 0 & 0 & 0 & 0 & 0 & 0 & 0 & 0 & 0 & 0 & \kappa _3 & 0 \\
 \kappa _2 & 0 & 0 & 0 & 0 & \kappa _2 & 0 & 0 & 0 & 0 & \kappa _2 & 0 & 0 & 0 & 0 &
   \kappa _1 \\
\end{array}
\right),  
\end{equation}
where,
$\kappa_1=\frac{1}{4} \left(Q_1+3 Q_4\right),\kappa_2=\frac{1}{4} \left(Q_1-Q_4\right)$ and $\kappa_3=\frac{1}{4} \left(-Q_1-4 Q_2-4 Q_3-3 Q_4+1\right)$. 

$\rho_{HL}^{2qq}$ is {\it not} in normal form, in which ``the Bloch representation of $\rho_{AB}$ would have 
$\vv{\textbf{a}}=0$ and $\vv{\textbf{b}}=0$, that is, the local density matrices would be maximally mixed''. In fact, the Bloch vectors of the two reduced $4 \times 4$ 
subsystems both have a component $\frac{1}{16} \sqrt{\frac{3}{2}} (Q_1+3 Q_4)$ associated with the fifteen generator 
of $SU(4)$. The components associated with the fourteen other generators are all zero in both cases. 
(A constructive way of bringing a single copy of a quantum state into normal form under local filtering operations 
was presented in \cite{verstraete2003normal}. A Matlab program for accomplishing this is given in \cite{filternormalform},) The Li-Qiao framework requires such normal forms.

The requirement that $\rho_{HL}^{2qq}$ is a nonnegative definite density matrix--or, equivalently, that its sixteen leading nested minors are nonnegative \cite{prussing1986principal}--takes the form \cite[eq. (29)]{slater2019bound}
\begin{equation} \label{d=4Basic}
Q_1>0\land Q_4>0\land Q_2>0\land Q_3>0\land Q_1+4 \left(Q_2+Q_3\right)+3 Q_4<1.
\end{equation}
The constraint that the partial transpose of $\rho_{HL}^{2qq}$ is nonnegative definite is \cite[eq. (30)]{slater2019bound}
\begin{equation} \label{d=4PPT}
Q_3>0\land Q_1+3 Q_4>0\land Q_1+4 \left(Q_2+Q_3\right)+3 Q_4<1\land Q_1^2+4 Q_2
   Q_1+Q_4^2  
\end{equation}
\begin{displaymath}
+16 Q_2 \left(Q_2+Q_3\right)+12 Q_2 Q_4<4 Q_2+2 Q_1 Q_4\land
   \left(Q_1-Q_4\right){}^2<16 Q_3^2 . 
\end{displaymath}
With these formulas, we are able to establish that the corresponding PPT-probability is $\frac{1}{2}+\frac{\log \left(2-\sqrt{3}\right)}{8 \sqrt{3}} \approx 0.404957$ (again, quite elegant, but seemingly of a different analytic form than the $d=3$ counterpart of $\frac{8 \pi}{27 \sqrt{3}}$). In \cite[sec. IIIB]{slater2019bound}, we obtained free entanglement and bound-entangled probability CCNR-based estimates of 0.4509440211445637 and 0.01265489845176, respectively.

Then, our 4-variable (as opposed to 32-variable [in Li-Qiao framework]) computations show that--if we maximize over simply the PPT states--we have $p=\frac{3^{24}}{2^{134}} \approx 1.2968528306 \cdot 10^{-29}$  (for $Q_1=\frac{3}{16},Q_2=\frac{9}{64},Q_3=\frac{3}{64},
Q_4=0$) and 
$s=\frac{49}{16} \approx 3.0625$ for the same four parameters. Now, if we exclude from the PPT states those that are bound-entangled according  to the realignment criterion, we obtain $s=\frac{9}{4} \approx 2.25$ 
(for $Q_1=0,Q_2=\frac{1}{4},Q_3=0,Q_4=0$), while $p$ appears to be unchanged. If we enforce the $p>\frac{3^{24}}{2^{134}}$ constraint, our estimate of the associated entanglement probability is 0.31711552, while the $s>\frac{9}{4}$ constraint gives us 0.39717107. 
Unfortunately, at this point in time, we do not have an {\it exact} entanglement probability--as in the two-qutrit case studied above--to which to fit the Li-Qiao entanglement constraint bounds.

Further analyses should be pursued in order to obtain the eight atoms spanning the 256-dimensional entanglement-probability three-set boolean algebra of the two-{\it ququart} Hiesmayr-L{\"o}ffler  magic simplex of Bell states. The main impediment, it seems, to doing so
is a lack of  precise knowledge as to the proper lower bound for the $P$ constraint--only knowing presently that   $\frac{3^{24}}{2^{134}} \approx 1.2968528306 \cdot 10^{-29}$ is greater than it, while we do know that $s=\frac{9}{4} \approx 2.25$ is the proper bound for the $S$ constraint. However, from the discussion in sec.~\ref{twoququart}, it would appear to be of interest to pursue an analysis employing $s=\frac{9}{4} \approx 2.25$ and $p=\frac{3^{24}}{2^{134}}$. 

In fact, such an attempt--based on 3,645,771 quasirandom four-dimensional points--yielded the eight atomic estimates,
\begin{equation} \label{estimates1}
\{2.7429040 \cdot 10^{-6},0.00108784,0.,0.314977,0.403877,0.0958316,0.,0.184224\},    
\end{equation}
where the same ordering of the atoms as indicated in (\ref{8atoms}) was employed.
We know already through symbolic integration that the PPT probability is $\frac{1}{2}+\frac{\log \left(2-\sqrt{3}\right)}{8 \sqrt{3}} \approx 0.404957$. We see that the estimate  for the fifth atom $\neg P \land \neg S \land PPT$ is quite close in value, that is, 0.404023. This atom corresponds to the separable states, so the estimate, in being slightly less than the 
PPT probability--due to the possibility of bound-entanglement--is plausible in that regard. If the lower bound for $p$ could be found, then, it seems reasonable that the three zero or near-zero estimates (all corresponding to atoms with $P$, rather than $\neg P$) would increase. Despite our lack of full knowledge as to the proper value of $p$ to employ, we can utilize our atomic estimates to obtain estimates {\it free} of $P$. For example, for the constraint $S$, just by itself,  the derived estimate--obtained by summing the first, second, fourth and sixth atomic estimates (\ref{estimates1})--is 
0.4118991565. Further, the derived estimate of $S \and PPT$, that is, 0.0010906, is close to $\frac{3}{2750} \approx 0.00109091$, and that of $S \lor PPT$, that is, 0.815776, is close to $\frac{31}{38} \approx 0.815789$.

If we limit our considerations to PPT-states for which $s \leq \frac{9}{4}$, the entanglement bound for $P$ appears to be at least as large as $ \frac{3^{24}}{2^{134}} \approx 8.50915 \cdot 10^{-31}$.

However, further numerical analysis suggested that the $P$ upper bound could be lowered--from $\frac{3^{24}}{2^{134}} \approx 1.2968528306 \cdot 10^{-29}$--to
$10^{-30}$ (for $Q_1=\frac{1}{5}, Q_2=\frac{1}{10}, Q_3=\frac{1}{20},Q_4=0$). To so improve our knowledge of the lower bound for $P$, we utilized our confidence in the full knowledge of the $S$ constraint, to eliminate states entangled according to that single criterion from further consideration. (However, though doing so might prove sufficient to fully determine the proper $P$ constraint--it is by no means clear that that is in fact the situation, seeing that it is not so in the two-qutrit case, as Table~\ref{tab:Second} indicates.)

Then, we were able--by finding some computational improvements--to increase our quasirandom point collection to size 101,215,383, now yielding the eight Hiesmayr-L{\"o}ffler two-ququart atomic 
estimates of 
\begin{equation} \label{estimates2}
\{0.000187037,0.000910652,0.,0.351977,0.40386,0.0588246,0.,0.18424\}.    
\end{equation}

Further investigation revealed a two-ququart PPT state ($Q_1=\frac{5}{806},Q_2=\frac{100}{407},Q_3=\frac{64}{36743},Q_4=\frac{8}{18805}$) with the apparently very small value $p \approx 1.553764401 \cdot 10^{-63}$, for which, nevertheless, $s \approx 2.2508113649 >\frac{9}{4}$, and is, thus, entangled. (So, the entanglement of this state would not be revealed--by higher settings for $p$--as seems not inconsistent with the Li-Qiao two-constraint [$P,S$] framework. Numerical fine-tuning reduces the indicated $p$ value further still to $4.86133\cdot 10^{-67}$.)
\section{Concluding Remarks}
In our analyses here, the CCNR (computable cross-norm realignment) criterion \cite{chen2002matrix,shang2018enhanced} for entanglement proves to be equivalent to the properly enforced constraint--involving the square of the Ky Fan norm (the sum of the singular values) \cite[eq. (32)]{li2018separable} of the correlation matrix in the Bloch representation--on $S$. Whether this equivalence is true, in general, is a question to be addressed.  (In certain auxiliary analyses, we  concluded that in the Hiesmayr-L{\"o}ffler  $d=3$ [two-qutrit] magic simplex model, the CCNR is equivalent--and not inferior, as can be the case \cite{shang2018enhanced}--to the  ESIC [SIC POVMs] test \cite{shang2018enhanced}, in yielding the same sets of entangled and bound-entangled states.
Efforts to similarly compare the CCNR and ESIC criteria  in the $d=4$ [two-ququart] version have so far proved too computationally challenging to complete.)

An outstanding problem is the conversion of the two-ququart Hiesmayr-L{\"o}ffler density matrix (\ref{d=4HL}) into normal form 
\cite{convert}. Although there are numerical approaches to this problem \cite{verstraete2003normal,filternormalform}, its symbolic character makes it still more challenging.

\begin{acknowledgements}
This research was supported by the National Science Foundation under Grant No. NSF PHY-1748958. I thank A. Mandilara for providing me with the Mathematica code by which I was able to corroborate the nature of the pseudo-one-copy undistillable states generated.
\end{acknowledgements}

\bibliography{main}

\end{document}